\documentclass[pra,aps,twocolumn,showpacs]{revtex4}

\usepackage{amssymb}
\usepackage{graphicx}

\usepackage{amsmath}

\def\<{\langle}
\def\>{\rangle}
\def\ket#1{|#1\>}
\def\bra#1{\<#1|}

\begin{document}

\title{Process fidelity estimation of linear optical quantum CZ gate: A comparative study }

\author{M. Mi\v{c}uda}
\affiliation{Department of Optics, Palack\'{y} University, 17. listopadu 1192/12, CZ-771 46 Olomouc, Czech Republic}

\author{M. Sedl\'{a}k}
\affiliation{Department of Optics, Palack\'{y} University, 17. listopadu 1192/12, CZ-771 46 Olomouc, Czech Republic}

\author{I. Straka}
\affiliation{Department of Optics, Palack\'{y} University, 17. listopadu 1192/12, CZ-771 46 Olomouc, Czech Republic}

\author{M. Mikov\'{a}}
\affiliation{Department of Optics, Palack\'{y} University, 17. listopadu 1192/12, CZ-771 46 Olomouc, Czech Republic}

\author{M. Du\v{s}ek}
\affiliation{Department of Optics, Palack\'{y} University, 17. listopadu 1192/12, CZ-771 46 Olomouc, Czech Republic}

\author{M. Je\v{z}ek}
\affiliation{Department of Optics, Palack\'{y} University, 17. listopadu 1192/12, CZ-771 46 Olomouc, Czech Republic}

\author{J. Fiur\'{a}\v{s}ek}
\affiliation{Department of Optics, Palack\'{y} University, 17. listopadu 1192/12, CZ-771 46 Olomouc, Czech Republic}

\begin{abstract}
 We present a systematic comparison of different methods of fidelity estimation of a linear optical quantum controlled-Z gate implemented by two-photon interference
 on a partially polarizing beam splitter. We have utilized a linear fidelity estimator based on the Monte Carlo sampling technique as well 
as a non-linear estimator based on maximum likelihood reconstruction of a full quantum process matrix. In addition, we have 
also evaluated lower bound on quantum gate fidelity determined by average quantum state fidelities for two mutually unbiased bases. 
In order to probe various regimes of operation of the gate we have introduced a tunable delay line between the two photons. 
This allowed us to move from high-fidelity operation to a regime where the photons become distinguishable and the success probability 
of the scheme significantly depends on input state. We discuss in detail possible systematic effects that could influence the 
 gate fidelity estimation. 
 \end{abstract}


\maketitle

\section{Introduction}

Most quantum computation and quantum information processing schemes rely on devices that transform quantum states 
while preserving their purity and quantum coherence. For example, in quantum circuit model of computation 
the elementary steps of the computation - quantum gates - are intended to be unitary transformations.
Ideally, the gates should operate deterministically according to a given prescription and the goal of experimentalist 
is to approach this regime as closely as possible. However, the experimentally implemented gates always somewhat 
deviate from the ideal ones due to various practical imperfections, 
thus creating some general transformation - a quantum channel. In some types of experiments, e.g. in quantum optics, 
the implementation of the gate may even be probabilistic and as a consequence the actually implemented transformation 
is a general probabilistic quantum operation. 

Motivated by the need to benchmark the experimentally implemented quantum gates and to identify their errors and imperfections, 
development of tools for experimental characterization of quantum operations has attracted considerable attention during recent years.
Several approaches have been proposed that differ in terms of the required resources as well as in the amount of information they provide. 
Often we want to understand precisely how the gate operates and we want to know exactly all its imperfections. 
Quantum process tomography  \cite{Poyatos97,Chuang97,Fiurasek01,Paris04} serves exactly this purpose and provides us with the full description of the gate 
for example in terms of its Choi operator $\chi$ \cite{Jamiolkowski72,Choi75,OBrien04b}. 
However, a complete quantum tomography requires resources which grow exponentially 
with the number of qubits unless one can assume that the Choi matrix $\chi$ has a small rank $r$ in which case one can apply compressed sensing \cite{Gross10,Shabani11}.
This motivated the search for other efficient methods of quantum gate characterization, whose goal is to determine only some specific 
features of the gate. Typically, we wonder how close is the actual gate to the ideal one and as a measure we use quantum process fidelity.

In 2005, it was shown by Hofmann that the quantum process fidelity can be efficiently bounded by measuring the
average quantum state fidelities for two mutually unbiased bases \cite{Hofmann05,Reich13}. 
This procedure has received a considerable attention
and it was utilized in several experiments to estimate the fidelity of a quantum CNOT gate \cite{Okamoto05,Bao07,Clark09,Gao10,Gao10b,Zhou11},  Toffoli gate \cite{Micuda13}, 
and multiqubit unitary operations on qubits carried by trapped ions \cite{Lanyon11}. 
If one wants to determine the exact value of the gate fidelity without performing full quantum state tomography, 
one can resort to Monte Carlo sampling techniques  \cite{Emerson07,Dankert09,Flammia11,Silva11,Steffen12}. The main advantage of Monte Carlo sampling 
is that the fidelity estimation error depends on the number of measurements and not on the size of the system which makes this approach particularly 
suitable for characterization of operations on large numbers of qubits. However, even for small-scale systems the Monte Carlo 
sampling may reduce the number of measurements below that required for full quantum process tomography.

In this paper we present a systematic comparison of different methods of  fidelity estimation of a linear optical quantum controlled-Z (CZ) gate. 
In the computational basis, this two-qubit gate introduces $\pi$ phase shift if and only if both qubits are in state $|1\rangle$, 
 \begin{equation}
U_{\mathrm{CZ}}=|00\rangle\langle 00|+|01\rangle\langle 01|+|10\rangle\langle 10|-|11\rangle\langle 11|. 
\label{VCZ}
\end{equation}
Recall that the CZ gate is equivalent to the CNOT gate up to single-qubit Hadamard transform on the target qubit.
We have performed full quantum process tomography of the gate and we have also estimated the quantum process fidelity by  Monte Carlo sampling 
and determined the Hofmann lower bound on the process fidelity. 
A peculiar feature of the linear optical quantum gates is that they are probabilistic \cite{Kok07}
hence generally they need to be described by trace decreasing quantum maps and the success probability of such gate may depend on the input state.
Recently, we have shown that the Hofmann bound is applicable to such probabilistic operations but the average state fidelities have to be calculated 
as weighted means with weights equal to the relative success probabilities for each input probe state \cite{Micuda13}.  
Here, we explicitly demonstrate that 
by using the ordinary state averages instead of the weighted ones one could actually overestimate the gate fidelity. Since the Hofmann bound 
has been applied in the past to characterize probabilistic linear optical CNOT gates in several experiments, 
we investigate in depth the influence of unequal success probabilities on the fidelity bounds.

For this purpose we deliberately introduce a tunable temporal delay between two photons whose polarization states represent the qubits on which the gate acts. By changing this delay
we can move from high-fidelity operation, where success probabilities for all input states are almost equal, to a regime where the photons  
become distinguishable and the success probabilities exhibit significant variations. We find that our experimental results are generally 
in agreement with theoretical expectations. Nevertheless, we observe certain minor differences between the fidelity estimates determined by
full process tomography and by Monte Carlo sampling, that are larger than statistical uncertainty. Also, in the high visibility regime we find 
that the Hofmann lower bound apparently slightly exceeds the 
estimated gate fidelity. We discuss possible systematic effects that could influence performance of the gate and explain these discrepancies.

The rest of the paper is organized as follows. The experimental setup is described in Section II. The quantum process fidelity estimation methods 
are reviewed in Section III. In Section IV we describe a simple theoretical model that shows how the fidelity of linear optical CZ gate depends on 
visibility of two-photon interference. Experimental results are presented and discussed in Section V. 
Finally, Section VI contains a brief summary and conclusions.

\section{Experimental setup}

We employ time correlated photon pairs generated in the process of frequency-degenerate  spontaneous 
parametric downconversion in a $2~\rm{mm}$ long BBO crystal cut for type II phase matching and pumped with $110$~mW continuous wave laser diode 
with central wavelength of $405~\rm{nm}$ \cite{Jezek11}. The orthogonally polarized signal and idler photons are spatially separated at a polarizing beam 
splitter (PBS), coupled into single mode fibers, and released back into free space at the input of the experimental setup shown in Fig.~1.
Qubits are encoded into polarization states of the photons and an arbitrary state of each qubit can be prepared  using a sequence of 
quarter-wave plate (QWP) and half-wave plate (HWP). Computational basis states are associated with horizontal
 and vertical polarization as $\ket{0}\equiv\ket{H}$, $\ket{1}\equiv\ket{V}$.  Besides the computational basis states we also use
  diagonally and anti-diagonally linearly polarized states 
\begin{equation}
|D\rangle=\frac{1}{\sqrt{2}}(|H\rangle+|V\rangle), \qquad |A\rangle=\frac{1}{\sqrt{2}}(|H\rangle-|V\rangle),
\end{equation}
as well as the left- and right-hand circularly polarized states
\begin{equation}
|R\rangle=\frac{1}{\sqrt{2}}(|H\rangle+i|V\rangle), \qquad |L\rangle=\frac{1}{\sqrt{2}}(|H\rangle-i|V\rangle).
\end{equation}

\begin{figure}[!t!]
\includegraphics[width=0.99\linewidth]{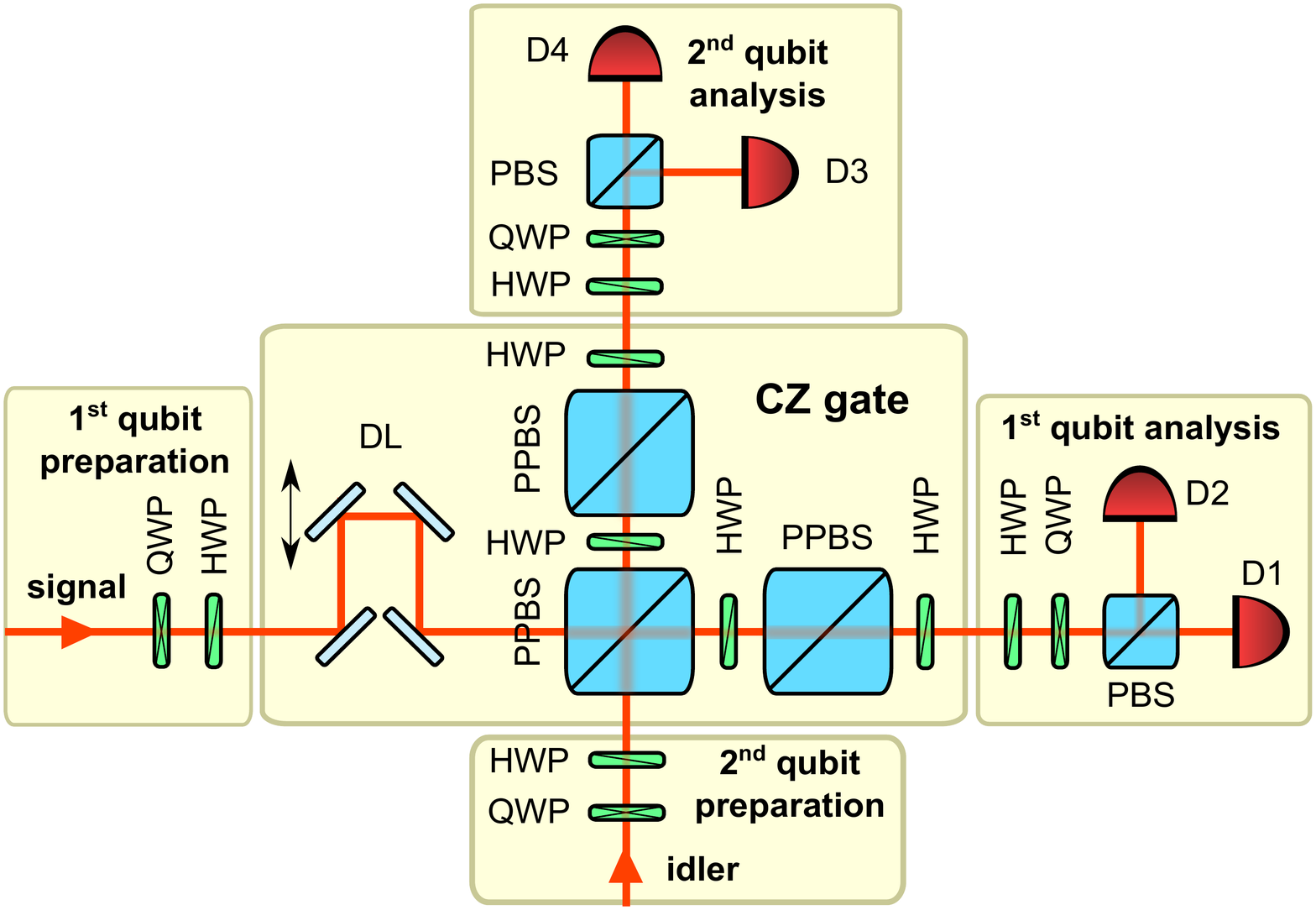}
\label{fig1}   
\caption{(Color online)
 Experimental setup. PPBS - partially polarizing beam splitter, PBS - polarizing beam splitter,
  HWP - half-wave plate, QWP - quarter-wave pate, D - single-photon detector, DL - tunable temporal delay line.}
\end{figure}

The quantum CZ gate  is implemented  by two-photon interference on a partially polarizing beam splitter (PPBS) that fully transmits horizontally  
polarized photons ($T_H=1$) while it partially reflects  vertically polarized photons ($T_V=1/3$) \cite{Ralph02,Hofmann02,Okamoto05,Langford05,Kiesel05,Lemr11}. 
The two-photon interference on the PPBS results in a $\pi$ phase shift if and only if both photons are vertically polarized, i.e. in logical state
 $|1\rangle$. 
The scheme also requires two additional PPBSs for balancing the amplitudes. 
Since all three partially polarizing beam splitters in our setup have the same splitting ratios, we use
additional half-wave plates 
rotated at $45^\circ$ to flip the horizontal and vertical polarizations. This ensures that the sequence of 
the central PPBS and the auxiliary PPBS acts as a polarization insensitive filter with effective transmittance $1/3$ for all polarizations.
 This linear optical gate operates in the coincidence basis \cite{Ralph02} 
 and its success is indicated by simultaneous detection of a single photon at each output port. 
 The gate is thus inherently conditional and its theoretical success probability reads $1/9$.

Polarization states of both output photons were analyzed by standard polarization measurement 
blocks consisting of half-wave plate, quarter-wave plate, polarizing beam splitter, and single-photon detectors. 
 In order to avoid the need to precisely calibrate relative detection efficiencies of the single-photon detectors, 
we have used only two-photon coincidences between single pair of detectors D2 and D3 for further data processing. 
Two-photon coincidences corresponding to measurement in any chosen product two-qubit basis 
were thus recorded sequentially and the measurement time of each number of coincidences was set to $30$~s.

\section{Fidelity estimation methods}

For our purposes, a quantum operation $\mathcal{E}$ is most conveniently described using the Choi-Jamiolkowski 
isomorphism \cite{Jamiolkowski72,Choi75}, that attributes to each completely positive map $\mathcal{E}$ 
a positive semidefinite operator $\chi$ on a tensor product of input and output Hilbert space. 
This operator can be intuitively defined as a density matrix of a quantum state obtained by applying the 
operation $\mathcal{E}$ to one part of a pure maximally entangled state $|\Phi^{+}\rangle$ on two copies of an input Hilbert space,
\begin{align}
\label{eq:defchoi}
\chi = \mathcal{I} \otimes \mathcal{E}(\Phi^{+}),
\end{align}
where $\mathcal{I}$ denotes the indentity operation, and 
$\Phi^{+}=|\Phi^{+}\rangle\langle \Phi^{+}|$ denotes a density matrix of pure state $|\Phi^{+}\rangle$. For two-qubit operations we explicitly have
\begin{equation}
|\Phi^{+}\rangle=\sum_{j,k=0}^1 |jk\rangle|jk\rangle.
\end{equation}
An input density matrix $\rho_{\mathrm{in}}$ is by $\mathcal{E}$ transformed into $\rho_{\mathrm{out}}=\mathcal{E}(\rho_{\mathrm{in}})$ 
which can be expressed as
  \begin{equation}
 \rho_{\mathrm{out}}
  =\mathrm{Tr}_{\mathrm{in}}[\rho_{\mathrm{in}}^T\otimes \mathbb{I}_{\mathrm{out}} \,\chi],
  \end{equation}
  where $T$ stands for the transposition in the computational basis and $\mathbb{I}$ denotes an identity operator.
For probabilistic operations, $\rho_{\mathrm{out}}$ is normalized so that its trace is equal to the success probability of $\mathcal{E}$ for input
state $\rho_{\mathrm{in}}$, 
\begin{equation}
p=\mathrm{Tr}[\rho_{\mathrm{in}}^T\otimes \mathbb{I}_{\mathrm{out}} \,\chi].
\end{equation}
The Choi matrix of a unitary CZ gate (\ref{VCZ}) reads
  \begin{align}
  \chi_{\mathrm{CZ}} &= (\mathbb{I}\otimes U_{\mathrm{CZ}}) \,|\Phi^{+}\rangle \langle \Phi^{+}| \,  (\mathbb{I}\otimes U_{\mathrm{CZ}}^\dagger) ,
  \label{chiCZ}
  \end{align}
  hence it is proportional to a density matrix of a pure maximally entangled state.
The process fidelity of quantum operation $\chi$ with respect to the unitary CZ gate is defined as a normalized overlap of their Choi matrices,
\begin{equation}
 F_\chi=\frac{\mathrm{Tr}[\chi \,\chi_{\mathrm{CZ}}]}{ \mathrm{Tr}[\chi_{\mathrm{CZ}}] \mathrm{Tr}[\chi] }.
 \end{equation}
 $F_\chi$ is sometimes called entanglement fidelity \cite{Horodecki99}, 
 because it is defined as an overlap of $\chi$ with a pure maximally entangled state.

\subsection{Quantum process tomography}

In our experiment, the CZ gate is probed with $36$ product two-qubit states $|\Psi_{jk}\rangle=|\psi_j\rangle|\psi_k\rangle$,
where the $6$ different single-qubit states $|\psi_j\rangle$ form three mutually unbiased bases,  
\begin{equation}
|\psi_j\rangle \in \{|H\rangle, |V\rangle,|D\rangle,|A\rangle,|R\rangle,|L\rangle\}.
\end{equation}
The measurements on the output two-photon states are products of single-qubit projective measurements, where each qubit is measured 
in one of the three bases $H/V$, $D/A$, and $R/L$, and we perform two-qubit measurements for all $9$ combinations of these bases.
The probability of projecting the output photons onto state $|\Psi_{lm}\rangle$ for input state $|\Psi_{jk}\rangle$ can be expressed as 
\begin{equation}
p_{jk,lm}=\mathrm{Tr} [\Psi_{jk}^T \otimes \Psi_{lm} \,\chi].
\end{equation}
The preparation of input probe states together with measurement on the output states can be interpreted as a quantum 
measurement on $\chi$, described by a POVM with $36\times 36$ elements
$\Pi_{jk,lm}= \Psi_{jk}^T \otimes \Psi_{lm}$. This POVM  satisfies the completeness relation 
\begin{equation}
\sum_{j,k=1}^6 \sum_{l,m=1}^6 \Pi_{jk,lm}= 81 \mathbb{I},
\label{POVMcomplete}
\end{equation}
and the  knowledge of all $p_{jk,lm}$ fully and unambiguously determines $\chi$.  Note that Eq. (\ref{POVMcomplete}) implies that 
\begin{equation}
 \sum_{j,k=1}^6 \sum_{l,m=1}^6  p_{jk,lm} = 81 \mathrm{Tr}[\chi].
 \label{psum}
\end{equation}

The measured coincidences $C_{jk,lm}$ exhibit Poissonian statistics with mean equal to $Np_{jk,lm}$, where $N$ is the average number of photon pairs 
generated by the source during the measurement time of $30$ s. 
We reconstruct the quantum operation $\chi$ from the measured coincidences with the help of maximum likelihood estimation \cite{Hradil97,Jezek03}.
The likelihood function representing the probability of measurement results $C_{jk,lm}$ for a given quantum operation $\chi$ can be expressed as 
\begin{equation}
\mathcal{L}= \prod_{j,k=1}^6 \prod_{l,m=1}^6 \frac{\left( Np_{jk,lm}\right)^{C_{jk,lm}}}{C_{jk,lm}!} e^{-N p_{jk,lm}}.
\end{equation}
It is convenient to work with the log-likelihood function $\ln \mathcal{L}$. The terms that do not depend on $\chi$ can be omitted and using Eq. (\ref{psum})
we obtain
\begin{equation}
\ln \mathcal{L}= \sum_{j,k=1}^6 \sum_{l,m=1}^6 {C_{jk,lm}} \ln p_{jk,lm} - \lambda \mathrm{Tr}[\chi],
\label{logL}
\end{equation}
where $\lambda=81N$. The actual pair generation rate $N$ is unknown due to various losses and imperfect photon collection and detection efficiency.
Therefore, $\mathrm{Tr}[\chi]$ can be effectively considered as a free parameter and we can set $\mathrm{Tr}[\chi]=1$
during the maximization of the log-likelihood function (\ref{logL}). Maximum likelihood estimation of probabilistic quantum operation then becomes
completely equivalent to quantum state estimation. The quantum operation $\chi$ which maximizes $\mathcal{L}$ satisfies the extremal equation \cite{Hradil97}
\begin{equation}
R \chi = \lambda \chi, 
\end{equation}
where 
\begin{equation}
R=\sum_{j,k=1}^6\sum_{l,m=1}^6 \frac{C_{jk,lm}}{p_{jk,lm}} \,  \Pi_{jk,lm},
\end{equation}
and the Lagrange multiplier $\lambda$ which fixes the trace of $\chi$ is proportional to the total number of coincidences, $\lambda=C_{\mathrm{tot}}/\mathrm{Tr}[\chi]$, where
\begin{equation}
C_{\mathrm{tot}}=\sum_{j,k=1}^6 \sum_{l,m=1}^6 C_{jk,lm}.
\end{equation}
The operation $\chi$ which maximizes $\mathcal{L}$ can be calculated by repeated iterations of symmetrized extremal equation, 
which preserves positive semidefiniteness of $\chi$ \cite{Jezek03},
\begin{equation}
\chi = \frac{R \chi R}{\mathrm{Tr}[R \chi R]}.
\end{equation}
As a starting point of the iterations we choose a full-rank operator $\chi_0=\mathbb{I}/16$, and the iterations 
are terminated when $|R\chi-\lambda \chi|_1 /C_{\mathrm{tot}} < 10^{-5}$, where $|A|_1=\sum_{j,k}|A_{jk}|$.

\begin{table}[t]
\caption{Non-zero coefficients $s_{abcd}$ in the expression (\ref{FchiMonteCarlo}) for process fidelity of quantum CZ gate.}
\label{table1}
\begin{ruledtabular}
\begin{tabular} {llllrcllllr}
$a$ & $b$ & $c$ & $d$ & $s_{abcd}$ & \qquad & $a$ & $b$ & $c$ & $d$ & $s_{abcd}$ \\
0 & 0 & 0 & 0 & 0.25 & & 2 & 0 & 2 & 3 & -0.25 \\
0 & 1 & 3 & 1 & 0.25 & & 2 & 1 & 1 & 2 & 0.25  \\
0 & 2 & 3 & 2 & -0.25 & & 2 & 2 & 1 & 1 & 0.25   \\
0 & 3 & 0 & 3 & 0.25 & & 2 & 3 & 2 & 0 & -0.25  \\
1 & 0 & 1 & 3 & 0.25 & & 3 & 0 & 3 & 0 & 0.25 \\
1 & 1 & 2 & 2 & 0.25 & & 3 & 1 & 0 & 1 & 0.25 \\
1 & 2 & 2 & 1 & 0.25 & & 3 & 2 & 0 & 2 & -0.25 \\
1 & 3 & 1 & 0 & 0.25 & & 3 & 3 & 3 & 3 & 0.25  \\
\end{tabular}
\end{ruledtabular}
\end{table}
  
\subsection{Monte Carlo sampling}

Here we review the estimation of quantum process fidelity by Monte Carlo sampling as proposed 
in Refs. \cite{Flammia11,Silva11} and we pay special attention to the fact
that we deal with probabilistic trace-decreasing operations. 
The operator $\chi_{\mathrm{CZ}}$ defined in Eq. (\ref{chiCZ}) can be expanded in the operator basis formed by tensor products of Pauli matrices,
\begin{equation}
\chi_{\mathrm{CZ}}= \sum_{a,b,c,d=0}^3  s_{abcd} \, \sigma_a\otimes\sigma_b \otimes \sigma_c \otimes \sigma_d.
\label{chiCZexpansion}
\end{equation}
It will be helpful to express the Pauli operators in terms of projectors onto the probe states $|\psi_j\rangle$,

\def\projector#1{|\makebox[9pt][c]{$#1$}\rangle\langle\makebox[9pt][c]{$#1$}|}
\begin{align}
\nonumber
\sigma_0 = \projector{H} + \projector{V},    \\\nonumber
\sigma_1 = \projector{D} - \projector{A} ,   \\\nonumber
\sigma_2 = \projector{R} - \projector{L},    \\
\sigma_3 = \projector{H} - \projector{V}.
\label{sigma}
\end{align}

 Due to the orthogonality relations $\mathrm{Tr}[\sigma_a\sigma_b]=2\delta_{ab}$, the coefficients 
in the expansion (\ref{chiCZexpansion}) can be determined as follows,
\begin{equation}
s_{abcd}= \frac{1}{16} \mathrm{Tr}[\chi_{\mathrm{CZ}} \, \sigma_a\otimes\sigma_b \otimes \sigma_c \otimes \sigma_d].
\label{sabcd}
\end{equation}
For CZ gate one finds that only 16 of the coefficients (\ref{sabcd}) are nonzero \cite{Steffen12} and these coefficients are listed in Table I. 
On inserting the expansion (\ref{chiCZexpansion}) into the formula for $F_\chi$ we obtain
\begin{equation}
F_\chi= \frac{1}{4 \mathrm{Tr}[\chi]} \sum_{a,b,c,d} s_{abcd} \mathrm{Tr}[\sigma_a\otimes\sigma_b \otimes \sigma_c \otimes \sigma_d \,\chi].
\label{FchiMonteCarlo}
\end{equation}
If we insert the expressions (\ref{sigma}) into Eq. (\ref{FchiMonteCarlo}) and make use of the identity 
(\ref{psum}),  we find that $F_\chi$ can be written as a ratio of linear functions of probabilities $p_{jk,lm}$,
\begin{equation}
F_{\chi}=\frac{81}{4}\frac{\sum_{j,k,l,m=1}^6 u_{jk,lm} p_{jk,lm}}{ \sum_{j,k,l,m=1}^6  p_{jk,lm}},
\label{FchiMCprobabilities}
\end{equation}
where the  coefficients $u_{jk,lm}$ are certain linear combinations of $s_{abcd}$. 
Note that the expression (\ref{FchiMCprobabilities}) for $F_\chi$ is not unique because the single-qubit identity operator $\sigma_0$ can be expressed 
in different ways in terms of the projectors onto $|\psi_j\rangle$.
For instance, instead of formula (\ref{sigma}) we can use  $\sigma_0=|D\rangle\langle D|+|A\rangle\langle A|$ or
$\sigma_0=|R\rangle\langle R|+|L\rangle\langle L|$. Also the normalization factor $\mathrm{Tr}[\chi]$ can be expressed 
in terms of the probabilities $p_{jk,lm}$ in many different ways. 
Since the (mean values of) the measured coincidences $C_{jk,lm}$ are proportional to $p_{jk,lm}$,
we can replace the probabilities with coincidences in Eq. (\ref{FchiMCprobabilities}) to obtain an estimator of the process fidelity,
\begin{equation}
F_{\mathrm{MC}}=\frac{81}{4}\frac{\sum_{j,k,l,m=1}^6 u_{jk,lm} C_{jk,lm}}{ \sum_{j,k,l,m=1}^6  C_{jk,lm}}.
\label{FchiMCcoincidences}
\end{equation}
Since we are able to collect enough data to estimate all terms in the expansion (\ref{FchiMCprobabilities}), 
we do not need to perform random sampling of only some of those terms as prescribed by the generic Monte Carlo sampling procedure \cite{Flammia11,Silva11}. 
Note, however, that such random sampling is  extremely useful for large systems, because it ensures that the total number 
of measurements that need to be carried out depends only on the required precision of fidelity estimation and not on the system size \cite{Flammia11,Silva11}.

\subsection{Hofmann bounds}

As shown by Hofmann \cite{Hofmann05}, a lower and upper bound on the process fidelity $F_\chi$ can be obtained from average state fidelities evaluated 
for two mutually unbiased bases. In case of CZ gate it is particularly suitable to use the product basis  $\{|DH\rangle,|DV\rangle,|AH\rangle,|AV\rangle\}$
and a dual  basis obtained  from the first basis by Hadamard transform on each qubit, $\{|HD\rangle,|VD\rangle,|HA\rangle,|VA\rangle\}$.
In what follows we shall label these bases as $1$ and $2$ and we denote by $|\omega_{j,k}\rangle$ a $j$th state of basis $k$.
The unitary CZ gate transforms all input states $|\omega_{j,k}\rangle$  onto output product states, 
\begin{equation}
\begin{array}{rclcrcl}
U_{\mathrm{CZ}}|DH\rangle&=&|DH\rangle, & \qquad &  U_{\mathrm{CZ}}|HD\rangle&=&|HD\rangle,   \\[1mm]
U_{\mathrm{CZ}}|DV\rangle&=&|AV\rangle, & \qquad &  U_{\mathrm{CZ}}|VD\rangle&=&|VA\rangle,   \\[1mm]
U_{\mathrm{CZ}}|AH\rangle&=&|AH\rangle, & \qquad &  U_{\mathrm{CZ}}|HA\rangle&=&|HA\rangle,   \\[1mm]
U_{\mathrm{CZ}}|AV\rangle&=&|DV\rangle, & \qquad &  U_{\mathrm{CZ}}|VA\rangle&=&|VD\rangle, 
\end{array}
\end{equation}
hence the state fidelities can be directly determined by measurements in product two-qubit bases.

The normalized output state of the quantum operation $\chi$ for the input $|\omega_{j,k}\rangle$ reads
\begin{equation}
\rho_{j,k}= \frac{1}{p_{j,k}} \mathrm{Tr}_{\mathrm{in}}[\omega_{j,k}^T\otimes \mathbb{I}_{\mathrm{out}} \chi],
\end{equation}
where $p_{j,k}=\mathrm{Tr}[\omega_{j,k}^T\otimes \mathbb{I}_{\mathrm{out}} \chi]$ is the success probability of $\chi$ for input $|\omega_{j,k}\rangle$
and $\omega_{j,k}=|\omega_{j,k}\rangle \langle \omega_{j,k}|$.  
The fidelity of the output state $\rho_{j,k}$ is defined as overlap with the pure state $U_{\mathrm{CZ}}|\omega_{j,k}\rangle$ produced by the unitary CZ gate, 
\begin{equation}
f_{j,k}= \langle \omega_{j,k} |U_{\mathrm{CZ}}^\dagger \,\rho_{j,k}\, U_{\mathrm{CZ}} |\omega_{j,k}\rangle.
\label{fjkdefinition}
\end{equation}
The average output state fidelity for $k$th basis is defined as a weighted mean of $f_{j,k}$ with weights equal to the 
success probabilities $p_{j,k}$ \cite{Bell12,Micuda13},
 \begin{equation}
 F_k= \frac{\sum_{j=1}^4 p_{j,k}f_{j,k}}{\sum_{j=1}^4 p_{j,k}}.
 \label{Fmeandefinition}
 \end{equation}
 Note that in order to determine $F_k$ we do not need  the absolute success probabilities but only the relative probabilities
 $P_{j,k}=p_{j,k}/\sum_{j'} p_{j',k}$. 
 
 Let $C_{j,j'}^k$ denote the number of coincidences corresponding to projections  
 onto a product state $U_{\mathrm{CZ}}|\omega_{j',k}\rangle$ for input probe state $|\omega_{j,k}\rangle$.  
 The state fidelities and relative success probabilities can be estimated as \cite{Micuda13}
 \begin{equation}
 f_{j,k}=\frac{C_{j,j}^k }{ S_{j}^k}, \qquad  P_{j,k}=\frac{S_j^k}{\sum_{j=1}^4 S_{j}^k},
 \end{equation}
 where $S_j^k=\sum_{j'=1}^4 C_{j,j'}^k$. On inserting these expressions into Eq. (\ref{Fmeandefinition}) we finally obtain
 \begin{equation}
 F_{k}= \frac{\sum_{j=1}^4 C_{j,j}^k}{\sum_{j=1}^4 S_{j}^k}.
 \label{Fmeancoincidences}
 \end{equation}
 In case of perfect gate operation only $C_{j,j}^k$ would be nonzero and $C_{j,j'}^k=0$ if $j\neq j'$. The average fidelity (\ref{Fmeancoincidences})
 is thus given by a ratio of the sum of the `good' coincidences $C_{j,j}^k$ and the sum of all the coincidences $C_{j,j'}^k$.
 
 Since $\sum_{j=1}^4 \omega_{j,k}=\mathbb{I}$ for all $k$, it holds that $\sum_{j=1}^4 p_{j,k}=\mathrm{Tr}[\chi]$
 and we can express the mean fidelities defined in Eq. (\ref{Fmeandefinition}) in a compact matrix form $F_k=\mathrm{Tr}[Q_k \chi]/\mathrm{Tr}[\chi]$, where
  \begin{equation}
  Q_k=\sum_{j=1}^{4} \omega_{j,k}^T\otimes \left(U_{\mathrm{CZ}}\omega_{j,k} U_{\mathrm{CZ}}^\dagger\right).
  \end{equation}
The gate fidelity $F_\chi$ can be bounded by the average state fidelities as follows \cite{Hofmann05,Micuda13}
\begin{equation}
 \max(F_1,F_2) \geq F_\chi\geq F_1+ F_2-1 \equiv F_H.
\label{Fbound}
\end{equation}
With the help of the above expressions one can rewrite the lower bound condition as
  \begin{equation}
  \frac{\mathrm{Tr}[Q\chi]}{\mathrm{Tr}[\chi]} \geq 0,
  \label{Finequality}
  \end{equation}
  where $Q= \frac{1}{4}\chi_{\mathrm{CZ}} -Q_1-Q_2+\mathbb{I}$.
  It can be shown by explicit calculation  that the operator $Q$ is positive semidefinite, 
  which proves that the inequality (\ref{Finequality}) holds for both deterministic and probabilistic quantum operations $\chi$ \cite{Micuda13}.

If all success probabilities $p_{j,k}$ are equal then the weighted means can be replaced by the ordinary means  $\bar{F}_k=\frac{1}{4}\sum_{j=1}^4 f_{j,k}$ 
and we obtain the Hofmann bound as originally formulated for deterministic operations, 
\begin{equation}
 F_\chi\geq \bar{F}_1+ \bar{F}_2-1 \equiv F_D.
\label{Fbounddet}
\end{equation}
We emphasize that this latter bound does not hold for probabilistic operations and $\bar{F}_1+\bar{F}_2-1$ may be larger than $F_\chi$
if $\chi$ is a trace decreasing map. In order to compare the two bounds (\ref{Fbound}) and (\ref{Fbounddet}) we write
\begin{equation}
 f_{j,k}=\bar{F}_k +\Delta f_{j,k}, \qquad p_{j,k}=\bar{p}+\Delta p_{j,k}, 
\end{equation}
 where $\bar{p}=\sum_j p_{j,k}/4 =\mathrm{Tr}[\chi]/4$. Since $\sum_j \Delta f_{j,k}=0$ and $\sum_j \Delta p_{j,k}=0$
by definition, we have
\begin{equation}
F_1+F_2= \bar{F}_1+\bar{F}_2+ \frac{1}{4 \bar{p}}\sum_{j=1}^4( \Delta p_{j,1} \Delta f_{j,1}+ \Delta p_{j,2} \Delta f_{j,2}).
\end{equation}
This formula reveals that the bounds (\ref{Fbound}) and (\ref{Fbounddet}) will differ considerably only if 
the state fidelities $f_{j,k}$ and success probabilities $p_{j,k}$ exhibit significant variations.

\section{Model of Linear optical CZ gate}

To experimentally probe various regimes of the gate operation including situation where the success probabilities significantly depend on the input states,
we deliberately introduce a variable time delay between the photons with the help of a delay line (DL), see Fig. 1. The time delay makes the photons partially 
or even fully distinguishable and it thus reduces the visibility of their interference \cite{Mikova13}.
In this section we theoretically analyze the impact of the reduced visibility of two-photon interference on the behavior of the gate. 
We will model this situation in a simple way: we assume that the two photons either interfere with probability $q$ or they behave 
as perfectly distinguishable particles with probability $1-q$. A more detailed model including 
also errors in transmittances of the partially polarizing beam splitters can be found in Ref. \cite{Nagata09}.

It is instructive to relate the value of the parameter $q$ to the visibility 
of Hong-Ou-Mandel (HOM) interference that can be directly measured experimentally. If we prepare signal photon in state $|V\rangle$, 
  idler photon in state $|H\rangle$, and set the waveplates in the $2$nd qubit analysis block such that it performs measurement in the $D/A$ basis,
  then a HOM dip can be observed by measuring the coincidences between detectors $D3$ and $D4$. The observed coincidence rate $C$ will be proportional to 
the photons' distinguishability, $C=C_\infty(1-q)$, where $C_\infty$ is the rate outside the dip. 
Visibility of two-photon interference  is defined as $\mathcal{V}=(C_\infty-C)/(C_\infty+C)$ and after some algebra  we obtain the relation
\begin{equation}
q=\frac{2\mathcal{V}}{1+\mathcal{V}}.
\label{relVq}
\end{equation}

The operation of the gate can be seen as a probabilistic mixture with probability $q$ of a perfect operation of the CZ gate 
(when the photons perfectly interfere and the gate succeeds with probability $1/9$) and of an incoherent transformation 
$\chi_{\mathrm{inc}}$ occurring otherwise. Thus, the Choi-Jamiolkowski operator corresponding to the gate reads
\begin{align}
\label{eq:chiformodel}
\chi= \frac{q}{9}\,\chi_{\mathrm{CZ}} + (1-q) \chi_{\mathrm{inc}}.
\end{align}
If the photons are distinguishable then 
 the gate operation still succeeds if both photons are either transmitted through or reflected from the central PPBS but these two contributions become 
 distinguishable and have to be added together incoherently.
After some algebra we thus find that $\chi_{\mathrm{inc}}$ is a mixture of an identity channel and an operation corresponding to projection onto state $|VV\rangle$,
\begin{equation}
\chi_{\mathrm{inc}}=\frac{1}{9}|\Phi^{+}\rangle \langle \Phi^+|+\frac{4}{9}\ket{VVVV}\bra{VVVV},
\end{equation}
 where we remind that
\begin{equation}
\ket{\Phi^{+}}=\ket{HHHH}+\ket{HVHV}+\ket{VHVH}+\ket{VVVV},
\end{equation}
in our current notation. The dependence of the gate fidelity  on visibility $\mathcal{V}$ can be determined using Eqs. (\ref{relVq})  and (\ref{eq:chiformodel}) 
and we get
\begin{align}
\label{eq:modelfid}
F_\chi=\frac{1+3\mathcal{V}}{4}.
\end{align}
We can see that the gate operates perfectly for $\mathcal{V}=1$ (or equivalently $q=1$) and has fidelity $1/4$ if we operate it out of the HOM interference ($\mathcal{V}=q=0$).

\begin{figure}[t]
\includegraphics[width=0.99\linewidth]{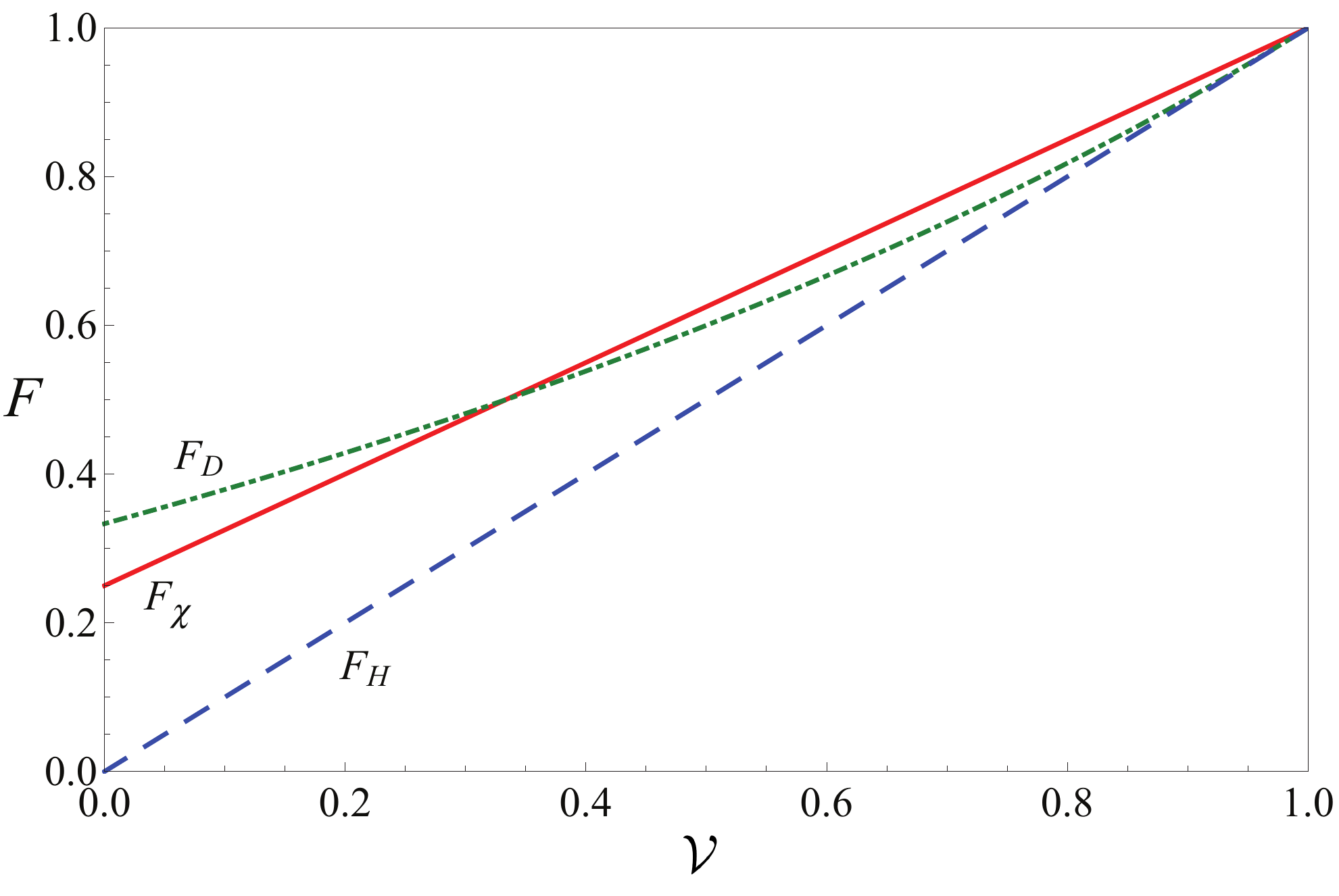}
\label{fig2}  
  \caption{(Color online) Dependence of the gate fidelity $F_{\chi}$ (solid red line), lower bound on gate fidelity $F_H$ (blue dashed line), 
  and lower bound valid for deterministic operations $F_D$ (green dot-dashed line) on two-photon interference visibility $\mathcal{V}$.}
\end{figure}

Let us now investigate the dependence of the Hofmann bound on the interference visibility. Our goal is to calculate 
the mean state fidelities $F_k$ and $\bar{F}_k$ and for this purpose we need to evaluate $p_{j,k}$ and $f_{j,k}$ as defined in Section III.C. 
It is convenient to rewrite the expression for success probability as $p_{j,k}=\mathrm{Tr}[\omega_{j,k}^T X]$, where
\begin{equation}
  \label{Xdef}
X=\mathrm{Tr}_{\mathrm{out}}[\chi]=\frac{1}{9}\mathbb{I}+\frac{4}{9}(1-q)\ket{VV}\bra{VV}.
\end{equation}
The four probe states $|HD\rangle$, $|HA\rangle$, $|DH\rangle$, $|AH\rangle$ have $\ket{0}\equiv\ket{H}$ as one of the qubits and the ideal 
CZ gate would act as an identity on them. Moreover, all these states are orthogonal to $|VV\rangle$. Therefore, we have for all these states
\begin{equation}
p_{j,k}=\frac{1}{9}, \qquad f_{j,k}=1,
\end{equation}
 irrespective of the value of visibility $\mathcal{V}$. On the other hand, the remaining four input states 
 $|V\!D\rangle$, $|V\!A\rangle$, $|DV\rangle$, $|AV\rangle$  have an overlap $1/\sqrt{2}$ with $\ket{VV}$ and we get
\begin{equation}
p_{j,k}=\frac{3-2q}{9}, \qquad  f_{j,k}=\frac{1}{3-2q},
\end{equation}
 for all of them.

\begin{figure*}[t]
\includegraphics[width=\linewidth]{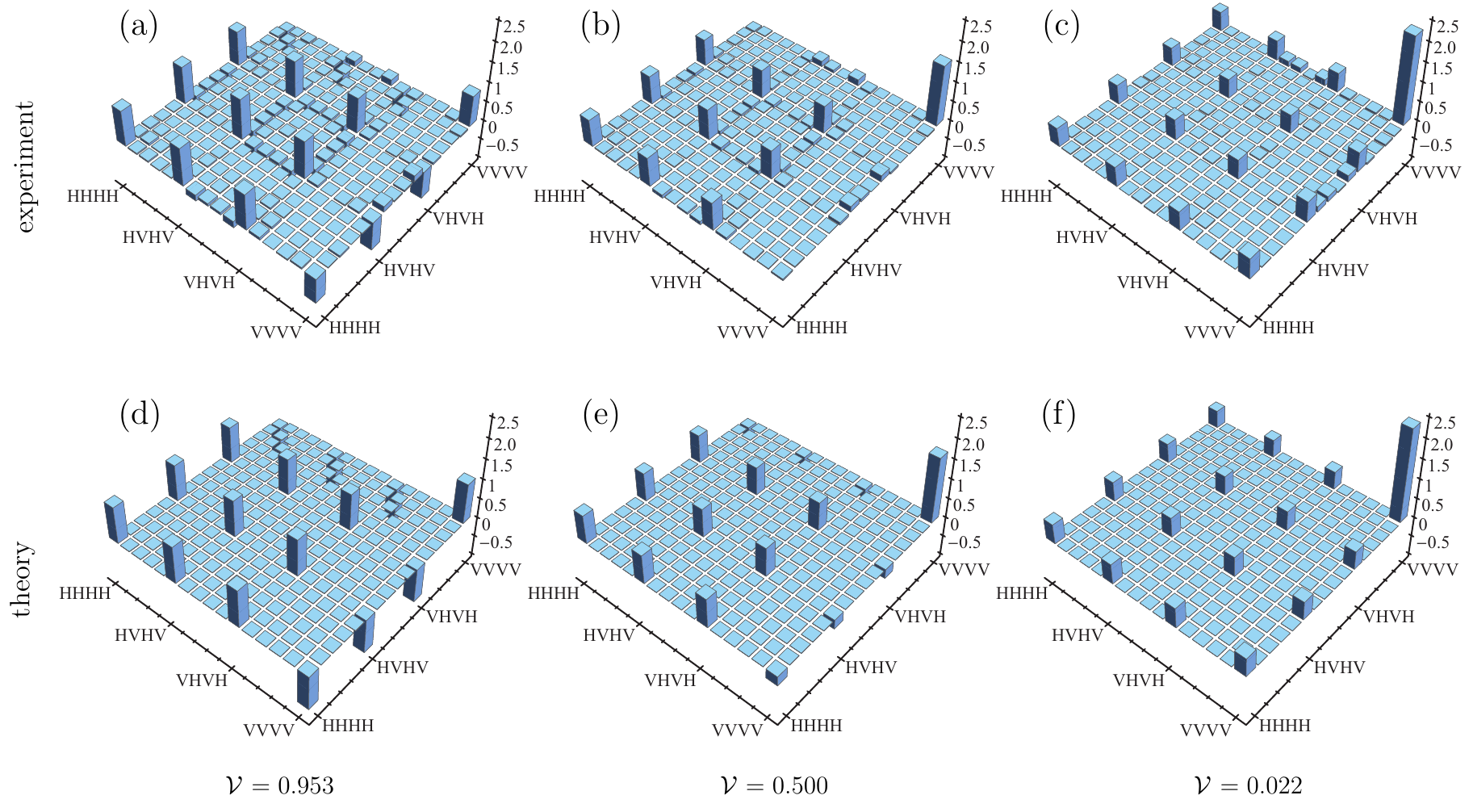}
\caption{(Color online) Quantum process matrices $\chi$ of linear optical CZ gate determined by Maximum likelihood reconstruction from experimental data (a,b,c), 
and theoretical process matrices (d,e,f) determined from the model presented in Sec. IV. The results are shown for three values of two-photon 
interference visibility $\mathcal{V}=0.953$ (a,d), $\mathcal{V}=0.50$ (b,e), and $\mathcal{V}=0.022$ (c,f). 
Imaginary parts of reconstructed $\chi$ represent a small noise background, and are not plotted.
To facilitate comparison, all matrices are normalized such that $\mathrm{Tr}[\chi]=4$.
}
\label{fig:chimatrices}
\end{figure*}

At this stage we are ready to evaluate $F_k$ and $\bar{F}_k$. Since each basis 
contains two states from the first and two states from the second above mentioned groups of states, we get
\begin{align}
F_1 =F_2=\frac{1+\mathcal{V}}{2}, \qquad \bar{F}_1 = \bar{F}_2 =\frac{2}{3-\mathcal{V}}.
\end{align}
The Hofmann bound (\ref{Fbound}) implies that the gate fidelity should satisfy
\begin{align}
 \mathcal{V} \leq F_{\chi} \leq \frac{1+\mathcal{V}}{2},
 \label{FchiinequalityV}
\end{align}
hence the lower bound on $F_\chi$ is directly equal to the visibility of two-photon interference $\mathcal{V}$. 
It is easy to see that the true fidelity (\ref{eq:modelfid}) indeed satisfies the inequalities (\ref{FchiinequalityV}) 
as it should be. In Fig. 2 we plot the true process fidelity as well as the Hofmann lower bound  in dependence on the interference visibility $\mathcal{V}$.

If we use the ordinary average state fidelities $\bar{F}_k$ instead of the weighted averages $F_k$, then we get
\begin{align}
\label{eq:ghffinal}
F_D \equiv \bar{F}_1+\bar{F}_2-1= \frac{1+\mathcal{V}}{3-\mathcal{V}},
\end{align}
which is larger than the true fidelity $F_\chi$ when $\mathcal{V}< \frac{1}{3}$, c.f. Fig. \ref{fig2}.
This explicitly demonstrates that the lower bound (\ref{Fbounddet}) is guaranteed to work only for deterministic operations
and its application to probabilistic operations may lead to overestimation of the process fidelity. 
Since $F_D-F_H=(1-\mathcal{V})^2/(3-\mathcal{V})$, the two bounds become very similar for high interference visibilities, 
and the difference becomes significant only for relatively low visibility, see Fig. 2.

\section{Results}

The tomographically complete measurements specified in Section III were performed for three different values of visibility of two-photon interference $\mathcal{V}$. 
The first measurement  was carried out at the Hong-Ou-Mandel dip where $\mathcal{V}=0.953$, which is the maximum visibility that we achieved with our setup. 
The second measurement was carried out with partly distinguishable photons ($\mathcal{V}=0.50$) and for the third measurement the temporal 
delay between the photons was increased such that they became completely distinguishable ($\mathcal{V}=0.022$).

The quantum process matrices determined by the Maximum Likelihood estimation procedure are plotted in Fig. 3. We can see that the shape 
of the reconstructed process matrices is in good agreement with the theoretical predictions for all the visibilities. In Table II 
we summarize the quantum process fidelities $F_{\chi}$ determined from the reconstructed quantum process matrices. 
The Table also contains process fidelities $F_{\mathrm{MC}}$ estimated by Monte Carlo sampling
and the Hofmann lower bound (\ref{Fbound}) on process fidelity $F_H$. For comparison, the table also includes a lower bound on process fidelity $F_D$ 
that is valid only for deterministic operations, c.f. Eq. (\ref{Fbounddet}). The coincidences and relative success probabilities required for evaluation of $F_H$ and $F_D$
are plotted in Fig. 4. The data are in good agreement with the prediction of the theoretical model described in Sec. IV. In particular, all the success probabilities are almost identical at the dip while
well outside the dip the states split into two groups whose success probabilities differ almost by a factor of $3$.

\begin{figure}[t]
 \centering
\includegraphics[width=\linewidth]{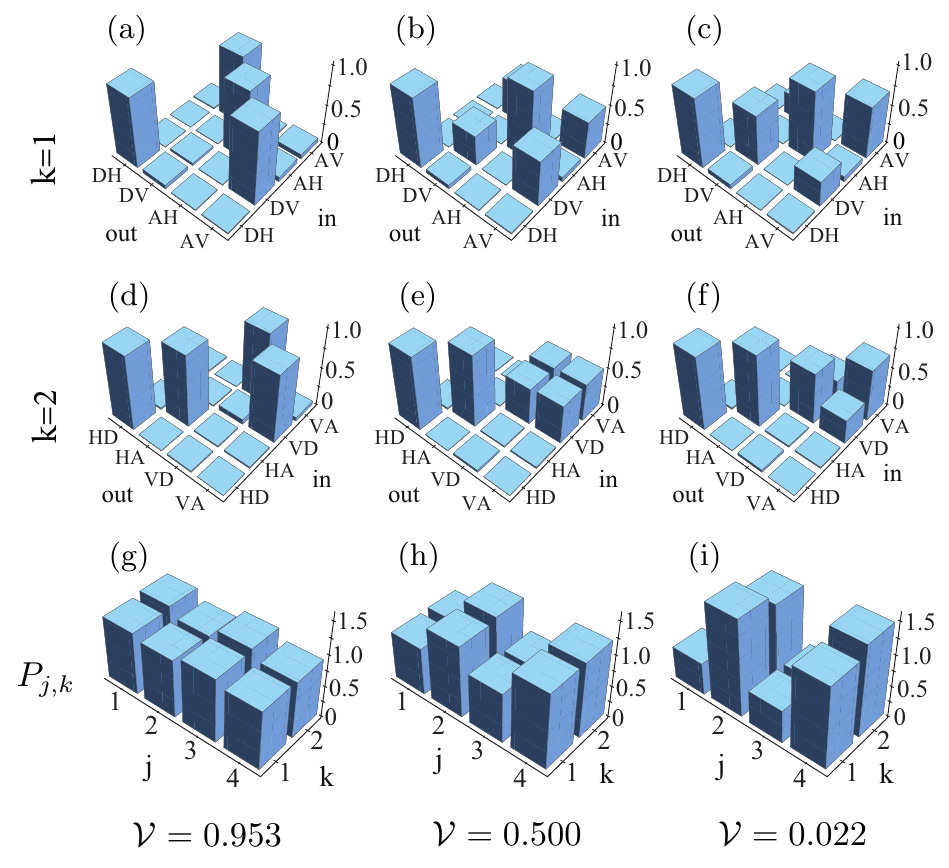}
\caption{(Color online) Normalized coincidences $C_{j,j'}^k/S_j^k$ (a)--(f) 
and relative success probabilities $P_{j,k}$ (g)--(i) used for determination of the Hofmann bound on quantum process fidelity.
The results are shown for the three visibilities $\mathcal{V}=0.953$ (a,d,g), $\mathcal{V}=0.500$ (b,e,h), and $\mathcal{V}=0.022$ (c,f,i).}

\label{fig:Hofmanndata}
\end{figure}

\begin{table}[b]
\caption{Experimentally determined quantum process fidelities $F_\chi$ and $F_{\mathrm{MC}}$, Hofmann lower bound on process fidelity $F_{H}$, 
lower bound  $F_D$ valid for deterministic operations, and upper bound on process fidelity provided by minimum of average state fidelities $F_1$ and $F_2$.
The results are shown for three values of visibility $\mathcal{V}$. }
\label{tableFidelities}
\begin{ruledtabular}
\begin{tabular}{cccccc}
 $\mathcal{V}$    &  $F_\mathrm{D}$ & $F_\mathrm{H}$ & $F_\chi$ & $F_{\mathrm{MC}}$ & $\mathrm{min}(F_1,F_2)$ \\\hline
 0.953  &    0.875(2) & 0.877(2) &  0.860(1)  &  0.871(2)   &  0.934(1)  \\
 0.500  &  0.465(2) &  0.372(2) &  0.531(1)  &     0.539(1)     &  0.676(1)  \\
 0.022  &  0.253(2) & -0.034(2)  &  0.232(1)  &  0.252(1)  &  0.479(1)  
\end{tabular}
\end{ruledtabular}
\end{table}

Statistical uncertainty of  $F_{\mathrm{MC}}$, $F_{H}$, and $F_{D}$ was estimated assuming Poissonian statistics of the 
measured coincidences and using standard error propagation. After some algebra we find that the statistical uncertainty 
of the Monte Carlo fidelity estimate $F_{\mathrm{MC}}$ can be expressed as
\begin{equation}
\left(\Delta F_{\mathrm{MC}}\right)^2=\frac{1}{C_{\mathrm{tot}}}\sum_{j,k,l,m=1}^6 \frac{C_{jk,lm}}{C_{\mathrm{tot}}} 
\left( \frac{81}{4} u_{jk,lm}-F_{\mathrm{MC}}\right)^2,
\end{equation}
and the statistical uncertainty of the Hofmann bounds is given by 
\begin{eqnarray}
\left (\Delta F_H\right )^2&=&\sum_{k=1}^2 \, \frac{F_{k}(1-F_{k})}{ \sum_{j=1}^4 S_j^k}, \nonumber \\
\left (\Delta F_D\right )^2&=&\frac{1}{16}\sum_{k=1}^2\sum_{j=1}^4 \frac{f_{j,k}(1-f_{j,k})}{S_j^k}.
\end{eqnarray}
In order to estimate the statistical uncertainty of fidelity $F_\chi$ determined from the reconstructed process matrix $\chi$, 
we have performed repeated simulations of the experiment followed by maximum likelihood reconstruction of the process matrix. 
For each $\mathcal{V}$ this procedure yielded an ensemble of $100$ reconstructed quantum process matrices and a corresponding ensemble of process fidelities, 
whose spread as quantified by one standard deviation was consistently lower than $10^{-3}$. 
The statistical uncertainty of $F_\chi$ indicated in Table II therefore represents a conservative upper bound.

\begin{table}[t]
\caption{Monte Carlo estimates of quantum process fidelity determined from the original and renormalized coincidences are listed
for the three considered values of interference visibility $\mathcal{V}$ and three different expansions of single-qubit identity operator $\sigma_0$
leading to different Monte Carlo estimators.}
\label{tableFMC}
\begin{ruledtabular}
\begin{tabular}{cccc}
 $\mathcal{V}$  &  $\sigma_0$ & $F_{\mathrm{MC}}$ & $\tilde{F}_{\mathrm{MC}}$  \\ \hline
0.953 & H/V &  0.871(2) & 0.861(2)  \\
0.953 & D/A &  0.882(2) & 0.870(2)  \\
0.953 & R/L &   0.833(1) & 0.846(1) \\

0.500 & H/V &  0.539(1) & 0.533(2) \\
0.500 & D/A &  0.521(1) & 0.518(2) \\
0.500 & R/L &  0.515(1) & 0.520(1) \\

0.022 & H/V &   0.252(1) & 0.240(1) \\
0.022 & D/A &   0.245(1) & 0.240(1) \\
0.022 & R/L &   0.242(1) & 0.235(1) 

\end{tabular}
\end{ruledtabular}
\end{table}

 The experimentally determined process fidelities 
$F_\chi$ and $F_{\mathrm{MC}}$ are somewhat smaller than the fidelity $(1+3V)/4$ predicted by the theoretical model.
This can be partly explained by the imperfections of the three partially polarizing beam splitters \cite{Nagata09} whose measured 
transmittances $T_{H1}=0.983$, $T_{V1}=0.348$, $T_{H2}=0.983$, $T_{V2}=0.344$, $T_{H3}=0.984$, $T_{V3}=0.324$
slightly differ from the ideal values $T_H=1$ and $T_V=1/3$. Note also that the differences between MaxLik and Monte Carlo 
estimates are larger than statistical uncertainty. Moreover, in the high-visibility regime $\mathcal{V}=0.95$
the Hofmann lower bound $F_H$ exceeds both $F_\chi$ and $F_{\mathrm{MC}}$ by an amount that is larger than 
the statistical error. All these features indicate influence of some effects that introduce systematic errors.
 To further investigate this aspect of our experiment, we have determined Monte Carlo estimates of the process 
 fidelity using three different estimators. These estimators were 
obtained following the procedure described in detail in Sec. IIIB, where the  single-qubit identity 
operator was expressed in three different ways as a sum of projectors, 
$\sigma_0=|H\rangle\langle H|+|V\rangle\langle V|$, $\sigma_0=|D\rangle\langle D|+|A\rangle\langle A|$, 
or $\sigma_0=|R\rangle\langle R|+|L\rangle\langle L|$. The results are summarized in Table III. 
We can see that the three estimators lead to fidelity estimates that differ by amounts exceeding 
the statistical uncertainty and the differences are largest in the high-visibility regime of operation.

\begin{figure}[t]
\includegraphics[width=0.99\linewidth]{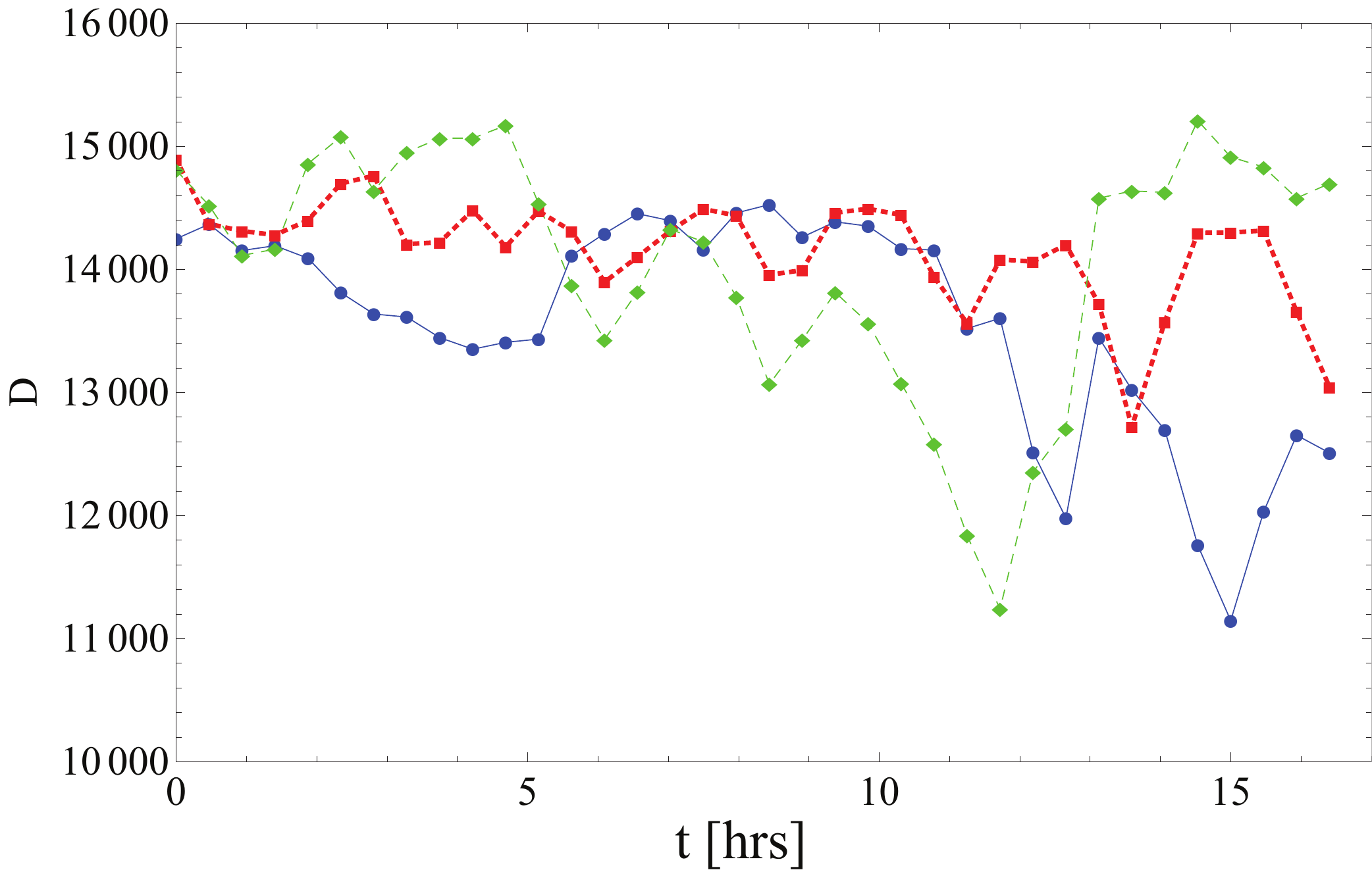}
  \caption{(Color online) 
The reference coincidences $D_{jk}$ are plotted as a function of time for the three considered two-photon visibilities $\mathcal{V}=0.953$ 
(solid blue line), $\mathcal{V}=0.50$ (red dotted line) and $\mathcal{V}=0.022$ (green dashed line). }
\label{fig4}
\end{figure}

Given the long duration of data acquisition (almost $17$ hours for each fixed $\mathcal{V}$), the measurements 
can be affected by long-term fluctuations of the rate of our source of correlated photon pairs. 
In order to track these fluctuations, we have performed additional coincidence measurements that 
can be used for data calibration. For each of the $36$ input states $|\Psi_{jk}\rangle$ we have measured the $36$
coincidences $C_{jk,lm}$ and then we have measured reference coincidences $D_{jk}$ for a fixed 
setting that did not depend on $j,k$ (input state $|HH\rangle$, projection onto $|HH\rangle$).
The dependence of the reference coincidences $D_{jk}$ on time is plotted in Fig. 5. The observed 
long term fluctuations are indeed non-negligible and should be accounted for in data processing. 
We therefore renormalize the measured coincidences, 
\begin{equation}
\tilde{C}_{jk,lm}= \frac{C_{jk,lm}}{D_{jk}}.
\label{Ccalibration}
\end{equation}
Monte Carlo  estimates of process fidelity $\tilde{F}_{\mathrm{MC}}$ obtained from the renormalized coincidences 
are listed in the last column of Table III. The data calibration leads to reduction of spread of the three estimates for each fixed visibility $\mathcal{V}$. 
The calibration (\ref{Ccalibration}) modifies the statistical uncertainty of the estimates, because 
the reference coincidences $D_{jk}$ are fluctuating quantities. Following the same procedure as before, we find that the statistical
error of $\tilde{F}_{\mathrm{MC}}$ is given by
\begin{eqnarray}
\left(\Delta \tilde{F}_{\mathrm{MC}}\right)^2=\frac{1}{\tilde{C}_{\mathrm{tot}}^2} \sum_{j,k,l,m=1}^6 
 \frac{\tilde{C}_{jk,lm}}{D_{j,k}} \left( \frac{81}{4}u_{jk,lm}-\tilde{F}_{\mathrm{MC}} \right)^2 \nonumber \\
+\frac{1}{\tilde{C}_{\mathrm{tot}}^2} \sum_{j,k=1}^6 \frac{1}{D_{jk}} 
\left[ \sum_{l,m=1}^6 \tilde{C}_{jk,lm}\left(\frac{81}{4}u_{jk,lm}-\tilde{F}_{\mathrm{MC}}\right)\right]^2. \nonumber \\
\end{eqnarray}
Explicit calculations reveal that the contribution due to fluctuations of $D_{jk}$ is almost negligible 
and the statistical uncertainty of $\tilde{F}_{\mathrm{MC}}$ is of the order of $10^{-3}$ similarly as for $F_{\mathrm{MC}}$, c.f. Table III.

Using the renormalized coincidences (\ref{Ccalibration}) we have also evaluated the Hofmann bounds $F_H$ and $F_D$ 
and the process fidelity $F_\chi$ obtained from the process matrix $\chi$ determined by maximum likelihood reconstruction.
It turns out that, in contrast to Monte Carlo sampling, the re-normalization has a negligible impact on these fidelity values. 
The largest difference occurs for $F_\chi$ at the high visibility regime ($\mathcal{V}=0.95$) where we get 
$F_\chi=0.860$ before renormalization and $F_\chi=0.858$ after renormalization. 
In all other cases, the difference between fidelities obtained from the original coincidences and the 
renormalized coincidences is smaller than $0.002$. Let us outline possible
 explanation of this robustness with respect to fluctuations  of pair generation rate. 
  Since all measurements for any given input state were performed in a row in a relatively short time span of cca $30$ minutes, 
 the long term fluctuations of the source rate have only a small impact on the estimation of state fidelities $f_{j,k}$ that 
 appear in expressions for $F_H$ and $F_D$. On the other hand, the maximum likelihood estimation combines together all the data 
 which form a significantly overcomplete set and therefore it in a sense averages over the long term fluctuations of the source.

 This analysis shows that the long-term fluctuations of the pair generation rate do not completely explain the observed discrepancies 
 between the fidelities. We therefore conclude that these discrepancies are caused by other systematic effects. 
 One such phenomenon could be a change of the visibility of two-photon interference during the measurement. 
 This is supported by the fact that the observed fidelity discrepancies are largest in the high-visibility regime. 
 In this case the setup is initially tuned to maximum visibility and thermal drifts and other effects cause reduction 
 of the visibility in the course of the measurement. By contrast, if the setup is operated with large temporal delay 
 between the two photons then small random changes of this delay do not have any impact on the
 performance of the scheme. Another possible source of systematic errors consists in imperfections of the wave plates
  and polarizing beam splitters that serve for state preparation and analysis.

\section{Conclusions}

In summary, we have compared several methods of quantum process fidelity estimation using the linear optical CZ gate as a suitable testing platform.
We have considered linear fidelity estimator based on the Monte Carlo sampling as well 
as a non-linear estimator based on maximum likelihood reconstruction of the full process matrix $\chi$. In addition, we have 
also evaluated lower bounds on quantum process fidelity provided by average quantum state fidelities. 
Since we have used the same data set to evaluate all the fidelities, the results admit direct comparison. 
We have observed good agreement between the Monte Carlo and MaxLik estimates and also the
fidelity bounds $F_{H}$ and $F_D$ behaved according to theoretical predictions. The observed small discrepancies 
between $F_{\mathrm{MC}}$ and $F_\chi$ can be partly attributed to fluctuations of the photon 
pair generation rate in the course of measurement, which were tracked by performing reference measurements 
and compenseted for by renormalization of the measured coincidences.

The remaining residual discrepancies between fidelity estimates can be attributed to various systematic effects 
such as change of the two-photon interference visibility during the measurement or small 
imperfections of the wave plates and polarizing beam splitters that are used for state preparation and analysis. 
In this context it is worth mentioning that it was shown very recently that fidelity estimation based on maximum 
likelihood reconstruction may lead to systematic underestimation of the fidelity \cite{Schwemmer13}. 
This underlies the importance of other more direct 
fidelity estimation techniques such as Monte Carlo sampling or fidelity bounds based on average state fidelities. 

By tuning the time delay between the two photons, we were able to control the visibility 
of two-photon interference and operate the gate in different regimes. In particular, when operated far outside the dip, 
the gate exhibits very low fidelity and significant dependence of success probability on the input state.
This flexibility allowed us to probe experimentally the influence of the varying 
success probabilities on the Hofmann lower bound on quantum process fidelity. For probabilistic gates, 
valid lower bound $F_H$ can be obtained with the help of weighted averages of state fidelities with weights represented 
by the relative success probabilities. In contrast, the bound based on ordinary averages of state fidelities 
is valid only for deterministic operations and may fail to provide a lower bound for probabilistic operations.
This is clearly demonstrated by our theoretical calculations and confirmed also by our experimental data. Well outside the dip we observe 
$F_{D}=0.253(2)$ while $F_{\chi}=0.232(1)$ and $\tilde{F}_{\mathrm{MC}}\leq 0.240(1)$. On the other hand, when the CZ gate 
is operated at the dip ($\mathcal{V}=0.95$), then the success probabilities are almost the same for all inputs, and $F_D$ and $F_H$ practically coincide. 
This confirms that the lower bounds on process fidelity of linear optical quantum gates reported in previous works \cite{Okamoto05,Bao07,Clark09,Gao10,Gao10b,Zhou11}
are reliable even if they were determined using ordinary averages of the state fidelities.

 \acknowledgments
 
 This work was supported by the Czech Science Foundation (project No. 13-20319S) and by Palacky University (project No. PrF-2013-008). 
 M.S. acknowledges support by the Operational Program Education for Competitiveness - European Social Fund 
 (project No. CZ.1.07/2.3.00/30.0004) of the Ministry of Education, Youth and Sports of the Czech Republic.


\begin{thebibliography}{99}

\bibitem{Poyatos97}
J.F. Poyatos, J.I. Cirac, and P. Zoller, Phys. Rev. Lett. \textbf{78}, 390 (1997).

\bibitem{Chuang97}
I.L. Chuang and M.A. Nielsen, J. Mod. Opt. \textbf{44}, 2455 (1997).

\bibitem{Fiurasek01}
J. Fiurasek and Z. Hradil, Phys. Rev. A \textbf{63}, 020101(R) (2001).

\bibitem{Paris04}
\emph{Quantum state estimation}, No. 649 in Lect. Notes Phys., M. Paris and J. \v{R}eh\'{a}\v{c}ek, eds., (Springer, Heidelberg, 2004).




\bibitem{Jamiolkowski72}
A. Jamiolkowski, Rep. Math. Phys. \textbf{3}, 275 (1972).

\bibitem{Choi75}
M.-D. Choi, Linear Algebra Appl. \textbf{10}, 285 (1975).


\bibitem{OBrien04b}
J.L. O'Brien, G.J. Pryde, A. Gilchrist, D.F.V. James, N.K. Langford, T.C. Ralph, and A.G. White, Phys. Rev. Lett. \textbf{93}, 080502 (2004).
 

\bibitem{Gross10}
D. Gross, Y.-K. Liu, S. T. Flammia, S. Becker, and J. Eisert, Phys. Rev. Lett. \textbf{105}, 150401 (2010).

\bibitem{Shabani11}
A. Shabani, R.L. Kosut, M. Mohseni, H. Rabitz, M.A. Broome, M.P. Almeida, A. Fedrizzi, and A.G. White, Phys. Rev. Lett. \textbf{106}, 100401 (2011).


\bibitem{Hofmann05}
H.F. Hofmann, Phys. Rev. Lett. \textbf{94}, 160504 (2005).


\bibitem{Reich13}
D. M. Reich, G. Gualdi, and C. P. Koch, Phys. Rev. A \textbf{88}, 042309 (2013).



\bibitem{Okamoto05}
R. Okamoto, H.F. Hofmann, S. Takeuchi, and K. Sasaki, Phys. Rev. Lett. \textbf{95}, 210506 (2005).

\bibitem{Bao07}
X.H. Bao, T.Y. Chen, Q. Zhang, J. Yang, H. Zhang, T. Yang, and J.W. Pan, Phys. Rev. Lett. \textbf{98}, 170502 (2007).


\bibitem{Clark09}
A.S. Clark, J. Fulconis, J.G. Rarity, W.J. Wadsworth, and J.L. O'Brien,  Phys. Rev. A \textbf{79}, 030303(R) (2009).

\bibitem{Gao10}
W.B. Gao, P. Xu, X.-C. Yao, O. G\"{u}hne, A. Cabello, C.-Y. Lu, C.-Z. Peng, Z.B. Chen, and J.W. Pan,  Phys. Rev. Lett. \textbf{104}, 020501 (2010).


\bibitem{Gao10b}
W.B. Gao, A.M. Goebel, C.Y. Lu, H.N. Dai, C. Wagenknecht, Q.A. Zhang, B. Zhao, C.Z. Peng, Z.B. Chen, Y.A. Chen, and J.W. Pan, PNAS \textbf{107}, 20869 (2010).



\bibitem{Zhou11}
X.Q. Zhou, T.C. Ralph, P. Kalasuwan, M. Zhang, A. Peruzzo, B.P. Lanyon, and J.L. O'Brien, Nature Commun. \textbf{2}, 413 (2011).



\bibitem{Micuda13}
M. Mi\v{c}uda, M. Sedl\'{a}k, I. Straka, M. Mikov\'{a}, M. Du\v{s}ek, M. Je\v{z}ek, and J. Fiur\'{a}\v{s}ek, 
 Phys. Rev. Lett. \textbf{111}, 160407 (2013).

 
\bibitem{Lanyon11}
B.P. Lanyon, C. Hempel, D. Nigg, M. M\"{u}ller, R. Gerritsma, F. Z\"{a}hringer, P. Schindler,  J.T. Barreiro, M. Rambach, G. Kirchmair, M. Hennrich,
P. Zoller, R. Blatt, and C.F. Roos,  Science \textbf{334}, 57 (2011).




\bibitem{Emerson07}
J. Emerson, M. Silva, O. Moussa, C. Ryan, M. Laforest, J. Baugh, D.G. Cory, R. Laflamme, Science \textbf{317}, 1893 (2007).


\bibitem{Dankert09}
C. Dankert, R. Cleve, J. Emerson, and E. Livine, Phys. Rev. A \textbf{80}, 012304 (2009).

\bibitem{Flammia11}
S. T. Flammia and Y.-K. Liu, Phys. Rev. Lett. \textbf{106}, 230501 (2011).

\bibitem{Silva11}
M. P. da Silva, O. Landon-Cardinal, and D. Poulin, Phys. Rev. Lett. \textbf{107}, 210404 (2011).


\bibitem{Steffen12}
L. Steffen, M. P. da Silva, A. Fedorov, M. Baur, and A. Wallraff, Phys. Rev. Lett. \textbf{108}, 260506 (2012).


\bibitem{Kok07}
P. Kok, W. J. Munro, Kae Nemoto, T. C. Ralph, Jonathan P. Dowling, and G. J. Milburn,
Rev. Mod. Phys. \textbf{79}, 135 (2007).


\bibitem{Jezek11}
M. Je\v{z}ek, I. Straka, M. Mi\v{c}uda, M. Du\v{s}ek, J. Fiur\'{a}\v{s}ek, and R. Filip, Phys. Rev. Lett. \textbf{107}, 213602 (2011).



\bibitem{Langford05}
N. K. Langford, T.J. Weinhold, R. Prevedel, K. J. Resch, A. Gilchrist, J. L. O'Brien, G. J. Pryde, and A. G. White,  Phys. Rev. Lett. \textbf{95}, 210504 (2005).

\bibitem{Kiesel05}
N. Kiesel, C. Schmid, U. Weber, R. Ursin, and H. Weinfurter, Phys. Rev. Lett. \textbf{95}, 210505 (2005).

\bibitem{Lemr11}
K. Lemr, A. \v{C}ernoch, J. Soubusta, K. Kieling, J. Eisert, and M. Du\v{s}ek, Phys. Rev. Lett. \textbf{106}, 013602 (2011).

\bibitem{Ralph02}
T. C. Ralph, N. K. Langford, T. B. Bell, and A. G. White, Phys. Rev. A \textbf{65}, 062324 (2002).


\bibitem{Hofmann02}
H.F. Hofmann and S. Takeuchi, Phys. Rev. A \textbf{66}, 024308 (2002).



\bibitem{Horodecki99}
M. Horodecki, P. Horodecki, and R. Horodecki, Phys. Rev. A \textbf{60}, 1888 (1999). 

    
\bibitem{Hradil97}
Z. Hradil, Phys. Rev. A \textbf{55}, R1561 (1997) 

\bibitem{Jezek03}
M. Je\v{z}ek, J. Fiur\'{a}\v{s}ek, and Z. Hradil  Phys. Rev. A \textbf{68}, 012305 (2003).




\bibitem{Bell12}
B. Bell, A.S. Clark, M.S. Tame, M. Halder, J. Fulconis, W.J. Wadsworth, and J.G. Rarity, New J. Phys. \textbf{14}, 023021 (2012).



\bibitem{Mikova13}
M. Mikov\'{a}, H. Fikerov\'{a}, I. Straka, M. Mi\v{c}uda, M. Je\v{z}ek, M. Du\v{s}ek, and R. Filip, Phys. Rev. A \textbf{87}, 042327 (2013).


\bibitem{Nagata09}
T. Nagata, R. Okamoto, H.F. Hofmann, and S. Takeuchi, New J. Phys. \textbf{12}, 043053 (2009).



\bibitem{Schwemmer13}
C. Schwemmer, L. Knips, D. Richart, T. Moroder, M. Kleinmann, O. Gühne, H. Weinfurter, arXiv:1310.8465 (2013).


\end{thebibliography}
\end{document}